\documentclass[aps,pra,preprint,superscriptaddress,amsmath,amssymb,amsfonts]{revtex4-1}

\usepackage{graphicx} % enhanced support for graphics
\usepackage{dcolumn}
\usepackage{pslatex} % use postscript fonts by default
\usepackage{bm}       % provides access to bold math symbols
\usepackage{amsmath}
\usepackage{textcomp} % allows upright mu
\usepackage{url}     % helps typeset URLs

\newcommand{\figurewidth}{0.7\columnwidth}
\graphicspath{ {./PaperFigures/} }

\begin{document}
\title{
Enhanced third-order optical nonlinearity driven by surface-plasmon field gradients
}

\author{Vasily Kravtsov}
\affiliation{Department of Physics, Department of Chemistry, and JILA, University of Colorado, Boulder, CO, 80309, USA}
\author{Sultan AlMutairi}
\affiliation{Department of Physics and Astronomy, Texas A\&M University, College Station, TX, 77843, USA}
\author{Ronald Ulbricht}
\affiliation{Department of Physics, Department of Chemistry, and JILA, University of Colorado, Boulder, CO, 80309, USA}
\author{A. Ryan Kutayiah}
\affiliation{Department of Physics and Astronomy, Texas A\&M University, College Station, TX, 77843, USA}
\author{Alexey Belyanin}
\affiliation{Department of Physics and Astronomy, Texas A\&M University, College Station, TX, 77843, USA}
\author{Markus B. Raschke}
\email{Corresponding author: markus.raschke@colorado.edu}
\affiliation{Department of Physics, Department of Chemistry, and JILA, University of Colorado, Boulder, CO, 80309, USA}
\date{\today}

\begin{abstract}

Achieving efficient nonlinear optical frequency conversion in small volumes is key for future on-chip photonic devices that would provide a higher-speed alternative to modern electronics.
However, the already intrinsically low conversion efficiency severely limits miniaturization to nanoscale dimensions. 
Here we demonstrate that gradient-field effects can provide for an efficient, conventionally dipole-forbidden nonlinear response, offering a new approach for enhanced nonlinear optics in nanostructures.
We show that a {\em longitudinal} nonlinear source current can dominate the third-order optical nonlinearity of the free electron response in gold in the technologically important near-IR frequency range where the nonlinearities due to other mechanisms are particularly small.
Using adiabatic nanofocusing to spatially confine the excitation fields, from measurements of the $2\omega_1 - \omega_2$ four-wave mixing response as a function of detuning $\omega_1 - \omega_2$, we find up to $10^{-5}$ conversion efficiency with a gradient field contribution to $\chi^{(3)}_{\mathrm{Au}}$ of up to $10^{-19}~\mathrm{m}^2 / \mathrm{V}^2$.
The results are in good agreement with theory based on plasma hydrodynamics.
Our results demonstrate an increase in nonlinear conversion efficiency with decreasing sample size that can offset and even overcompensate the volume decrease of conventional dipolar pathways.
This will enable more efficient nonlinear optical devices and frequency converters and facilitate the extension of coherent multidimensional spectroscopies to the nanoscale.

\end{abstract}

%\pacs{...}

\maketitle

%%%%%%%%%%%%%%%%%%%%%%%%%%%%%%%%%%%%%%%%%%%%%%%%%%%%%%%%
\section{Introduction}
\noindent
Nonlinear optics provides a platform for optical frequency conversion and all-optical information processing that can potentially overcome speed limitations of modern electronics and enable faster computing and data communication.
Device miniaturization and on-chip integration thus require an efficient nonlinear optical resonse in deep sub-wavelength volumes.
However, intrinsic optical nonlinearities are generally weak, which severely limits corresponding signal levels generated in nanoscopic regions.

Various approaches to achieve enhanced nonlinear conversion efficiencies have been considered, from exploring materials with high intrinsic nonlinearities~\cite{Moss2013} to the precise design of extrinsic driving field distribution~\cite{Segal2015}.
Simultaneous mode volume compression and local field engineering for enhanced nonlinear optics can be achieved through the use of photonic crystals~\cite{Lepeshkin2004}, metamaterials~\cite{Lapine2014}, microcavities~\cite{Lin1994}, structures with reduced speed of light~\cite{Monat2010}, or plasmon-resonant metallic nanoparticles providing nanometer-scale confinement and multifold field enhancement~\cite{Kauranen2012}.

A significant gain in efficiency in the third-order nonlinear response could be of particular interest as these optical processes allow, for example, all-optical switching and manipulation of ultrafast laser pulses through the intensity-dependent refractive index~\cite{MacDonald2009}.
Furthermore, four-wave mixing (FWM) is the basis for important ultrafast spectroscopy techniques, such as coherent multidimensional optical spectroscopy~\cite{Jonas2003} and coherent anti-Stokes Raman spectroscopy (CARS), which can be implemented with nanoscale resolution~\cite{Kravtsov2016} and single-molecule sensitivity using near-fields of plasmon-resonant particles~\cite{Zhang2014, Yampolsky2014}.
However, the discussion of near-field effects in the third-order optical nonlinearity has been limited to local field enhancement factors associated with particle plasmon resonances~\cite{Danckwerts2007, Kim2008, Palomba2009, Jung2009, Genevet2010}, with limited attention given to finite size effects~\cite{Hache1986, Hache1988, Lysenko2016}, surface contribution~\cite{Palomba2008}, and ponderomotive terms~\cite{Ginzburg2010}.

%AB
Here we demonstrate a different mechanism of the third-order optical nonlinearity in metallic nanoparticles or nanoantennas based on large longitudinal field gradients associated with strongly confined plasmonic fields. 
While radiation of the associated longitudinal nonlinear currents into propagating transverse far-field waves is forbidden for translationally invariant bulk metal, this restriction is relaxed on the nanoscale. 
We find that in subwavelength structures the gradient field mechanism becomes an efficient and even dominant source term in FWM. 
 %AB
Using plasmonic nanofocusing to spatially confine the excitation fields, we show that the corresponding third-order nonlinear optical susceptibility $\chi^{(3)}$ in the near-IR spectral range can be enhanced $> 5$ times when transitioning from non-degenerate to degenerate FWM with up to $10^{-5}$ conversion efficiency and a corresponding contribution to $\chi^{(3)}_{\mathrm{Au}}$ of up to $10^{-19}~\mathrm{m}^2 / \mathrm{V}^2$.
We further develop a simple plasma hydrodynamic model that provides excellent semi-quantitative description of the nonlinear conversion efficiency and underlying electron dynamics.

%AB
We emphasize that the demonstrated mechanism becomes efficient in metal structures smaller than the wavelength of light; it does not rely on size quantization of the electron motion which would be important only in nanoparticles of much smaller size (smaller than the electron de Broglie wavelength \cite{Hache1988}). 
Therefore, the field gradient effect is particularly attractive for long-wavelength applications, especially in the spectral range far from any electron resonances where other nonlinearity mechanisms become increasingly inefficient. 
It should also be universal for any nanostructure based on materials with large free carrier densities, including metals, semiconductors, or doped graphene. 
%AB

Our results contribute to the understanding of the fundamental mechanisms of optical nonlinearities in plasmonic nanoparticles and nanoantennas in general and suggest a new route for enhanced nonlinear optical conversion in small volumes.
This can find potential applications from all-optical on-chip nonlinear devices to multidimensional ultrafast spectroscopies with nanoscale spatial resolution.

\begin{figure}[htb]
	\includegraphics[width=\figurewidth]{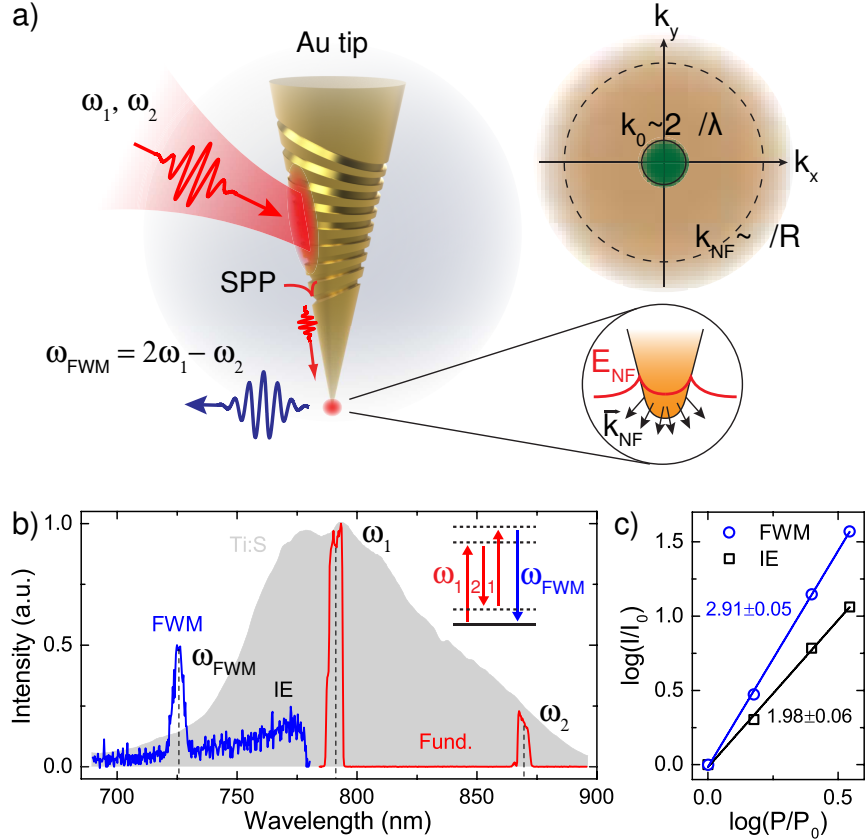}
	\caption{(a) Experimental configuration of grating-coupling and nanofocusing of femtosecond pulses for the generation of a nano-confined FWM response (left), with enhanced efficiency due to broad distribution of field momenta $k_{x,y}$ for nanofocused SPPs associated with gradient field and spatial confinement (right). (b) Pulse shaping of the laser spectrum into narrow-band excitations at frequencies $\omega_1$ and $\omega_2$, with full laser spectrum (gray), fundamental scattering off the tip apex (red), and blue-shifted response that consists of an incoherent emission (IE) background, and a coherent FWM peak (blue). (c) Excitation power dependence of the FWM signal (blue) and the incoherent emission (black) showing cubic and quadratic dependence, respectively.}
	\label{fig:Intro}
\end{figure}
%\\\\\\\\\\\\\\\\\\\\\\\\\\\\\\\\\\\\\\\\\\\\\\\\\\\\\\\\

\section{Experiment}
In the experiment, we measure the FWM signal generated at the apex of a Au tip with $\sim$10~nm radius as a generalized model element for nano-confined nonlinear conversion.
In order to spatially confine the excitation fields and eliminate background we use adiabatic nanofocusing and spatially filtered tip apex emission detection as shown in Fig.~\ref{fig:Intro}a.
Incident light is grating-coupled into surface plasmon polaritons (SPPs) that propagate toward the tip apex, adiabatically compress with accompanying field enhancement, and generate a nonlinear optical response predominantly in the nanoscopic apex volume~\cite{Berweger2011}.

For the FWM experiments, excitation is provided by two spectrally narrow bands of center frequencies $\omega_1$ and $\omega_2$ and full width at half maximum (FWHM) of 10~nm, derived from the output of a Ti:Sapphire laser producing 10~fs pulses at the center wavelength of 800~nm (Femtolasers), with full spectrum shown in Fig.~\ref{fig:Intro}b (gray).
Spectral phase and amplitude are controlled by a $4f$ pulse shaper based on a liquid crystal spatial light modulator (CRi SLM640).
We ensure a flat spectral phase across the full bandwidth of the optical field at the tip apex by performing multiphoton intrapulse interference phase scan (MIIPS) algorithm based on apex-generated second harmonic generation as feedback signal~\cite{Berweger2011}.
The two spectral bands at $\omega_1$ and $\omega_2$ are then obtained by blocking parts of the original spectrum with a tunable hardware amplitude mask in the Fourier plane of the pulse shaper.

The emission from the tip apex is spatially filtered to suppress grating-scattered light and detected spectrally resolved with a spectrometer and a liquid-nitrogen-cooled CCD camera (Princeton Instruments).
The fundamental scattering and FWM signal can be detected interchangeably by using either neutral density (OD = 5) or short pass (OD = 7) filters in the detection path.

For comparison, we perform a FWM measurement on a flat surface of a single-crystalline Au sample in far-field by focusing the pump radiation (50X, NA = 0.5, Olympus microscope objective) under normal incidence and using back-reflection detection geometry.

\section{Results}
Fig.~\ref{fig:Intro}b shows the tip apex-emitted spectrum: the fundamental SPP scattering with the two excitation bands $\omega_1$ and $\omega_2$ at long wavelengths ($>780$~nm, red) and the broad inelastic scattering of the SPPs and the sharp FWM signal peak at short wavelengths ($<780$~nm, blue).

The broad background is due to incoherent emission (IE) derived from electronic excitations within the \textit{sp}-band of Au~\cite{Huang2014}.
Its power dependence is close to quadratic ($1.98\pm 0.06$) for the spectrally integrated signal as shown in Fig.~\ref{fig:Intro}c (black), and exhibits a frequency-dependent power law exponent due to varying spectral shape (see Supplement) in agreement with previous studies~\cite{Huang2014, Haug2015}.

%
%\\\\\\\\\\\\\\\\\\\\\\\\\\\\\\\\\\\\\\\\\\\\\\\\\\\\\\\\
\begin{figure}[b]
	\includegraphics[width=\figurewidth]{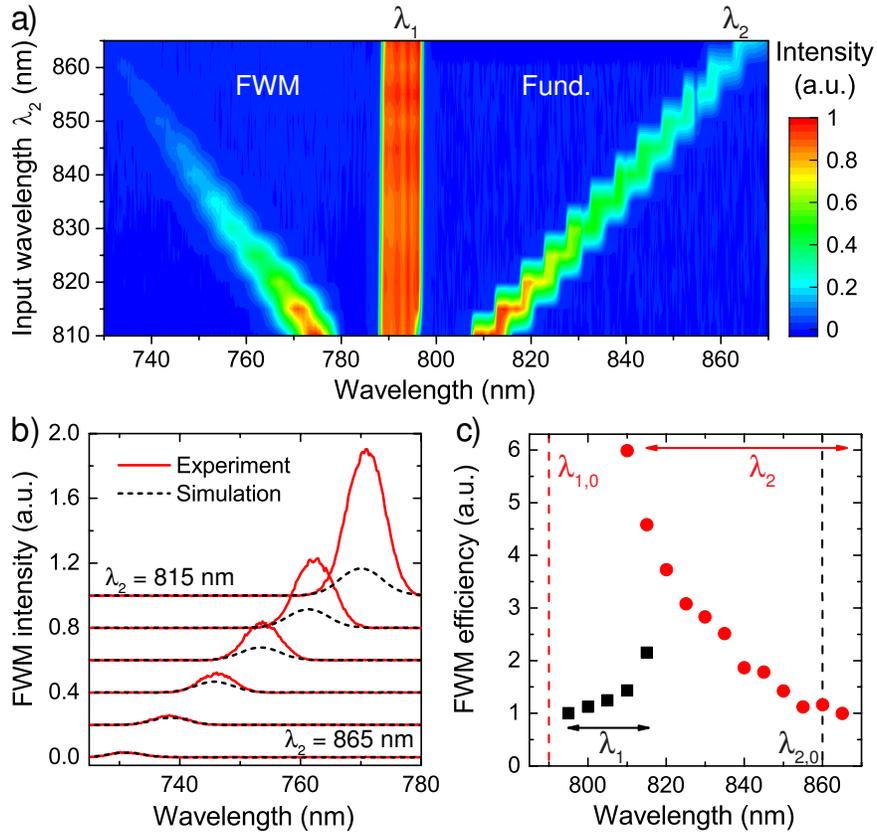}
	\caption{Measurement of the FWM efficiency: (a) FWM (left) and fundamental (right) tip-scattered spectra collected for varying excitation wavelength, (b) experimental FWM spectra for selected excitation wavelengths (red solid, $\lambda_1 = 790$~nm, $\lambda_2 =$ 815, 825, 835, 845, 855, and $865$~nm) together with calculated reference signal (black dashed), (c) FWM efficiency extracted in two experiments where either $\lambda_{1,0}$ (red) or $\lambda_{2,0}$ (black) are held constant, while varying $\lambda_2$ and $\lambda_1$, respectively.}
	\label{fig:Data}
\end{figure}
%\\\\\\\\\\\\\\\\\\\\\\\\\\\\\\\\\\\\\\\\\\\\\\\\\\\\\\\\
%

The coherent nonlinear FWM signal is centered at $\omega_\mathrm{FWM} = 2\omega_1 - \omega_2$.
Its power dependence is close to cubic (power law exponent $2.91 \pm 0.05$) as shown in Fig.~\ref{fig:Intro}c (blue), and depends quadratically and linearly on the incident intensities at $\omega_1$ and $\omega_2$, respectively, as expected.
With the estimated field enhancement of $\sim 25$ and the resulting peak electric field at the apex of $\sim 5\cdot 10^9$~V/m, the nano-FWM conversion efficiency~\cite{Kravtsov2016} reaches $\sim 10^{-5}$.

We then measure the spectral dependence of the efficiency of the FWM response.
As shown in Fig.~\ref{fig:Data}a, we collect FWM (left) and fundamental (right) spectra emitted from the tip apex for excitation with constant frequency $\omega_1$ while varying $\omega_2$.
Reference FWM spectra are calculated from integration over the fundamental scattering spectrum at the tip apex $\mathrm{I}(\omega)$, assuming a flat spectral phase and frequency-independent third-order susceptibility
\begin{equation}
\mathrm{I}_{\mathrm{ref}}(\omega ) \propto \biggl|\iiint\limits_{-\infty}^{\infty} \mathrm{d}\omega_{1,2,3}\, \biggr. \biggl. \sqrt{\mathrm{I}(\omega_1)\mathrm{I}(\omega_2)\mathrm{I}(\omega_3)}\, \delta(\omega - \omega_1 + \omega_2 - \omega_3)\biggr|^2,
\label{eq:FWM}
\end{equation}
and plotted in Fig.~\ref{fig:Data}b (black dashed) together with the experimental data (red solid) for selected values of $\omega_2$.
While peak position and overall spectral shape are consistent as expected, the measured FWM intensity increasingly exceeds reference values calculated from Eq.~\ref{eq:FWM} as $\omega_2$ approaches $\omega_1$.
We quantify that increase by calculating the FWM efficiency as a ratio between the experimental and reference signals at the peak maxima $\eta_\mathrm{FWM} = \mathrm{I}_\mathrm{FWM} / \mathrm{I}_\mathrm{ref}$.
This is plotted in Fig.~\ref{fig:Data}c (red circles) as a function of the varied excitation wavelength $\lambda_2$, and exhibits a 6-fold increase in the FWM efficiency within the spectral range of the measurement.
We then perform a corresponding measurement with variable $\lambda_1$ for fixed $\lambda_2$ (black squares) exhibiting an apparent opposite trend.
%
%\\\\\\\\\\\\\\\\\\\\\\\\\\\\\\\\\\\\\\\\\\\\\\\\\\\\\\\\
\begin{figure}[hb]
	\includegraphics[width=\figurewidth]{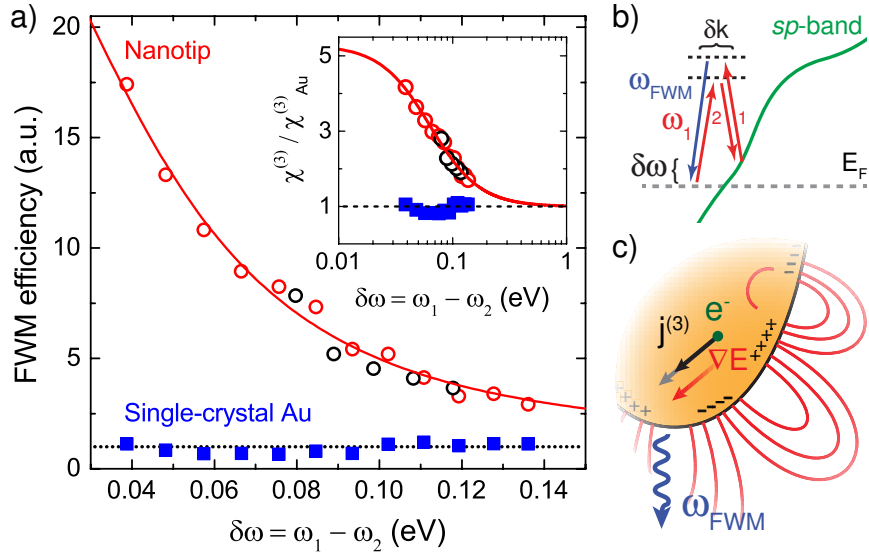}
	\caption{Modeling the FWM efficiency: (a) experimental result for nano-tip (red and black open circles) and single-crystalline Au surface in far-field (filled blue squares, dotted line is a guide to the eye), together with the model fit (red). Inset shows the enhancement of the third-order susceptibility of the tip relative to that of bulk Au as a function of detuning between the two excitation frequencies. (b) Schematic of the intraband FWM process. (c) Illustration of the FWM generated by longitudinal nonlinear currents and out-coupled into transverse fields.}
	\label{fig:Results}
\end{figure}
%\\\\\\\\\\\\\\\\\\\\\\\\\\\\\\\\\\\\\\\\\\\\\\\\\\\\\\\\
%
However, when plotting both datasets against the detuning $\delta\omega = \omega_1 - \omega_2$ as shown in Fig.~\ref{fig:Results}a, the FWM efficiency shows a universal increase with decreasing $\delta\omega$ for both measurements (red and black open symbols, left scale).
This indicates that the FWM mechanism is independent of the excitation frequency, and only depends on the detuning.
We note that a similar behavior of the FWM is observed for Au nanorods (see Supplement), although in that case the plasmon-resonant response and associated spectral variation of the local pump fields need additional consideration.
In contrast, the corresponding FWM result for a flat single-crystalline Au surface measured in far-field is shown in Fig.~\ref{fig:Results}a (filled blue squares), with low and independent of the detuning $\delta\omega$ efficiency.

\section{Discussion and conclusions}
In the following, we discuss FWM yield and its fundamentally distinct spectral behavior when generated in Au nanotip vs. single-crystalline Au surface, and the underlying enhancement mechanism.

In general, the third-order polarization $P^{(3)}(\omega_3)$ can be frequency-dependent through $\chi^{(3)}(\omega)$, local field enhancement factors at excitation and FWM frequencies $L(\omega), L(\omega_3)$, and excitation spectrum $E(\omega)$ with $P^{(3)}(\omega_3) \sim L(\omega_3)\chi^{(3)}(\omega) L^3(\omega)E^3(\omega)$.
In our case, all factors corresponding to the local fields at the excitation frequencies are contained in both measured $\mathrm{I}_\mathrm{FWM}$ and calculated reference $\mathrm{I}_\mathrm{ref}$ FWM intensities due to the spatially localized apex detection, and therefore cancel out for the calculated FWM efficiency $\eta_\mathrm{FWM} = \mathrm{I}_\mathrm{FWM} / \mathrm{I}_\mathrm{ref}$.
Then the experimental FWM efficiency spectrum will depend only on the local field factor at the FWM frequency and the nonlinear susceptibility of the material: $\eta_\mathrm{FWM} \propto \left[L(\omega_3)\chi^{(3)}(\omega_3; \omega_1, \omega_1, -\omega_2)\right]^2$.
Further, due to the off-resonant response of our tips around 800~nm, we can consider the local field factor $L(\omega_\mathrm{FWM})$ to vary only weakly across the narrow experimental range of FWM frequencies.
Therefore the spectral shape of the FWM efficiency is expected to mainly follow the spectral behavior of the nonlinear susceptibility itself $\eta_\mathrm{FWM} \propto \left[\chi^{(3)}(\omega; \omega_1, \omega_1, -\omega_2)\right]^2$.

For bulk Au, $\chi^{(3)}$ is weak in general with limited contributions from hot-electron, intraband, and interband terms~\cite{Boyd2014}.
The hot-electron term only becomes significant for pulse durations comparable to or exceeding the electron gas thermalization time of $\sim 500$~fs, and for excitation wavelength close to the \textit{d}- to \textit{sp}-band transition~\cite{Rotenberg2007, Boyd2014}.
In our case the FWM is generated by $< 100$~fs laser pulses away from the interband transition, so the hot-electron contribution is expected to be negligible.
Further, the \textit{sp}-band in Au is very close to parabolic, which implies zero restoring force and therefore vanishing intraband nonlinearity in the dipole approximation~\cite{Boyd2014}.
Additional higher-order magnetic-dipole and electric-quadrupole contributions to the nonlinear polarization are longitudinal and do not out-couple into transverse radiating modes.
For excitation with $\sim 1.56$~eV photon energies, the dominant contribution to the nonlinearity of bulk Au is expected to involve either two-photon resonant or one-photon off-resonant \textit{interband} electronic transitions between \textit{d}- and \textit{sp}-bands of Au~\cite{Kravtsov2016}.
The observed weak dependence of the FWM efficiency on the detuning $\delta\omega$ in bulk Au shown in Fig.~\ref{fig:Results}a (blue squares) is then consistent with the absence of resonant behavior in $\delta\omega$ for such transitions.

In nanostructures, in contrast, the \textit{intraband} contribution can become an efficient and leading source term.
Spatially compressed SPP modes at the tip apex $E = E_0 e^{i\mathbf{k_\mathrm{NF}}\cdot\mathbf{r}}$ exhibit strong field gradients $\left|\partial E / \partial\mathbf{r}\right| \propto k_\mathrm{NF} E$ in both transverse and longitudinal directions, corresponding to large linear momenta
 $p_\mathrm{NF}=\hbar k_\mathrm{NF} \sim \pi\hbar / R$, where $R$ is the tip apex radius.
For typical radii of $R\sim 10$~nm the near-field momenta can reach beyond 
%AB
$k_\mathrm{NF}\sim 3\cdot 10^6$~cm$^{-1}$, 
%AB
exceeding the corresponding far-field value for the given wavelength range by 2 orders of magnitude.
These in-plane momenta then allow for resonant and phase-matched electronic transitions within the \textit{sp}-band as illustrated in Fig.~\ref{fig:Results}b.
The dominant component of the third-order nonlinear current ${\bf j^{(3)}}$ is then generated by the associated \textit{longitudinal} electric fields of the plasmon modes.
The nanotip or nanoparticle geometry facilitates out-coupling of the longitudinal current  oscillations and resulting polarization density to the outgoing transverse electromagnetic waves as pictured in Fig.~\ref{fig:Results}c.
This is in fundamental contrast to the case of bulk Au where the corresponding radiative process is forbidden.

To provide a simple model to quantify the nonlinear response of the free electrons in a nano-structure (see Supplement for the full derivation), we consider the equation of motion for a degenerate electron plasma against the neutralizing background of immobile ions, neglecting the thermal motion, given by
\begin{equation}
\dfrac{\partial \mathbf{v}}{\partial t} + \gamma \cdot \mathbf{v} + \left( \mathbf{v} \cdot \nabla \right) \mathbf{v} -  \dfrac{e}{m_e} \mathbf{E}   -  \dfrac{e}{m_e c}  \mathbf{v} \times\mathbf{B} =0,
\label{eq:EqMotion}
\end{equation}
with external electric and magnetic fields $\bf{E}$ and $\bf{B}$, electron velocity $\bf{v}$, effective mass $m_e$, electron charge $e$, and collision rate $\gamma$.
We then consider longitudinal components of the velocity perturbation, fields, and their gradients along the $x$ axis,  $\mathbf{E} = E \hat{x}$  and $ \mathbf{v}=\mathrm{v} \hat{x} $, and neglect the contribution from the magnetic field.
We treat the electric field as a perturbation, approximate field gradients $\partial_{x}$ by $1/R$, and find a solution to Eq.~\ref{eq:EqMotion} in the form of a perturbation expansion.
For the third-order nonlinear term in the electron velocity $\mathrm{v}^{(3)}$ at frequency $\omega_3 = 2 \omega_1 - \omega_2$ to the two pump fields $E_1$ and $E_2$ at frequencies $\omega_1$ and $\omega_2$ we obtain:
\begin{equation}
\mathrm{v}^{(3)}\left(t\right) =     \dfrac{6e^{3}\gamma}{m_e^{3}R^{2}} \dfrac{E^{2}_{1}E^{*}_{2} e^{-i\omega_{3}t}}{\left(\gamma - i\omega_{1}\right)^{2}\left(\gamma + i\omega_{2}\right)\left(\gamma - i \omega_{3}\right)\left( \gamma^2 + \left( \omega_{1}-\omega_{2}\right)^2 \right)}.
\label{12}
\end{equation}

The third order susceptibility is related to the velocity through the nonlinear polarization  $P^{(3)}= \chi^{(3)} E_{1}^{2} E_{2}^{*}e^{-i  \omega_{3}t}$ and current density $j^{(3)} =\dfrac{dP^{(3)}}{dt}=en^{(0)}\mathrm{v}^{(3)}$, and can be expressed as
\begin{equation}
\chi^{(3)}_\mathrm{intra}= i \dfrac{6 n^{(0)}e^{4}}{\omega_{3}m_e^{3}R^{2}} \dfrac{ \gamma}{\left(\gamma - i\omega_{1}\right)^{2}\left(\gamma + i\omega_{2}\right)\left(\gamma - i \omega_{3}\right)\left( \gamma^2 + {\delta\omega}^2 \right)},
\label{eq:Lorentzian}
\end{equation}
where $n^{(0)}$ is the electron density.

This expression exhibits a resonance at $\omega_1 = \omega_2$ with a Lorentzian FWHM equal to $2\gamma$.
A fit to our experimental FWM efficiency $\eta_\mathrm{FWM}(\delta\omega) \propto \left(\chi^{(3)}_\mathrm{Au} + \chi^{(3)}_\mathrm{intra}(\delta\omega)\right)^2$, with $\gamma$ and $\chi^{(3)}_\mathrm{intra} (0)$ as the only free parameters, provides excellent agreement (Fig.~\ref{fig:Results}a, red solid line) with the experimental data.
The fit yields a FWHM of $128 \pm 14$~meV corresponding to the electron collision rate of $\gamma = 64 \pm 7$~meV, or mean free time between collisions of $\tau = 10.3 \pm 1.2$~fs, which is in good agreement with the Drude relaxation time of $\tau_\mathrm{D} = 9-14$~fs for Au~\cite{Johnson1972, Olmon2012}.

The second parameter we extract from the fit is the ratio of the intraband contribution at zero detuning to the third-order susceptibility of bulk Au with $\chi^{(3)}_\mathrm{intra} (0) / \chi^{(3)}_\mathrm{Au} = 4.3$.
The relative contribution of the gradient-induced nonlinearity is shown in the inset of Fig.~\ref{fig:Results}a, with the black dashed line indicating the intrinsic third order susceptibility of Au as extracted from the far-field measurement on the single-crystal Au sample.
We then estimate the absolute value of the third-order nonlinear susceptibility assuming all frequencies $\omega_{1,2,3} \sim 1.5$~eV, carrier density $n^{(0)} \sim 6\times 10^{22}$ cm$^{-3}$, $m_e$ equal to the effective electron mass in Au, and tip apex diameter $2R \sim 15$~nm:
\begin{equation}
|\chi^{(3)}| \sim \frac{6 n^{(0)}e^4}{m_e\gamma R^2 \omega_1^2 \omega_2 \omega_3^2} \sim 7.6 \times 10^{-12} \; {\rm esu}. 
\end{equation}
This corresponds to $|\chi^{(3)}| \sim 1.1\times 10^{-19}$~m$^2$/V$^2$ in SI units, and agrees with a previously reported value for Au of $2\cdot 10^{-19}$~m$^2$/V$^2$~\cite{Renger2010} for similar excitation conditions.

%\\\\\\\\\\\\\\\\\\\\\\\\\\\\\\\\\\\\\\\\\\\\\\\\\\\\\\\\
\begin{figure}[t]
	\includegraphics[width=\figurewidth]{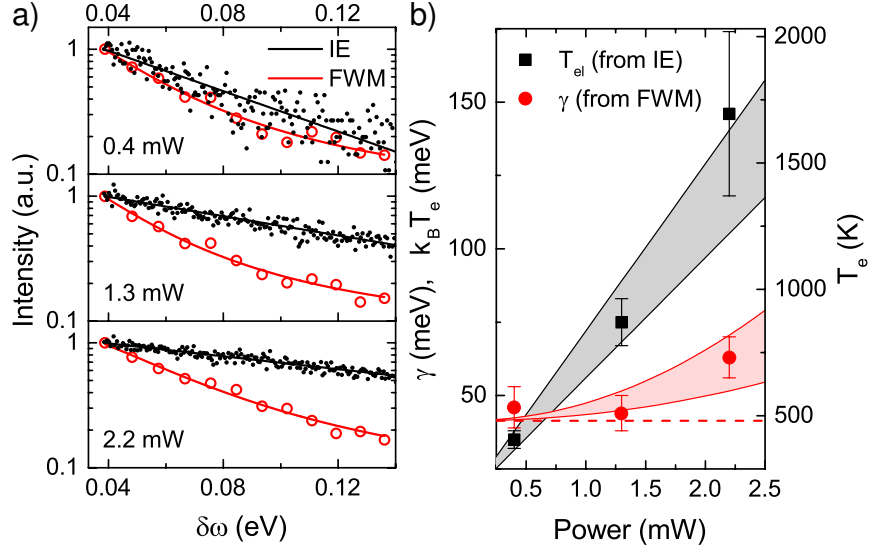}
	\caption{(a) FWM efficiency (red) and incoherent emission (IE) background spectra (black) for excitation power, varying from $0.4$~mW (top panel), to $1.3$~mW (middle), and to $2.2$~mW (bottom). (b) Electron gas temperature $T_e$ (black squares) extracted from IE fits, and electron collision rate $\gamma$ (red circles) extracted from FWM fits, for varying excitation power.}
	\label{fig:Power}
\end{figure}
%\\\\\\\\\\\\\\\\\\\\\\\\\\\\\\\\\\\\\\\\\\\\\\\\\\\\\\\\

Further, we measure several datasets of the tip-generated FWM efficiency vs. detuning for variable excitation power.
The results are summarized in Fig.~\ref{fig:Power}a, where the FWM efficiency (red circles) and corresponding Lorentzian fits to Eq.~\ref{eq:Lorentzian} (red lines) are plotted together with simultaneously acquired spectra of the incoherent emission (black dots) for the total power incident on the grating of 0.4~mW (top), 1.3~mW (middle), and 2.2~mW (bottom).
Although the exact mechanism of the incoherent emission in Au is still under debate~\cite{Haug2015, Hugall2015, Lin2016, Mertens2017}, it generally involves electronic transitions within the \textit{sp}-band, and therefore its spectrum depends sensitively on the temperature-dependent shape of the Fermi-Dirac distribution for the electron gas.
Our measured incoherent emission spectra therefore undergo a significant transformation with increasing incident power.
In agreement with previous work~\cite{Huang2014,Haug2015,Hugall2015}, we can approximately describe the spectra by a Boltzmann distribution $I_\mathrm{IE}(\delta\omega) \propto e^{-\delta\omega / k_B T_e}$, with the electron temperature $T_e$ increasing from $410 \pm 30$~K to $1700 \pm 300$~K for excitation powers between $0.4$~mW and $2.2$~mW.
The extracted electron temperature is shown in Fig.~\ref{fig:Power}b (black squares) together with an error-weighted linear fit, where the gray area indicates the uncertainty of the fit.

In contrast, the spectral shape of the FWM signal changes only slightly, which is consistent with our model that predicts the FWM dependence on the detuning to be fully defined by only (i) spatial extent of the field gradients and (ii) electron collision rate $\gamma$.
The former is defined by the tip apex geometry and therefore does not change with excitation power.
The latter is only weakly dependent on temperature for the range of laser intensities used in our experiment.
The extracted values of $\gamma$ are shown in Fig.~\ref{fig:Power}b (red circles) and exhibit a slight increase from $46 \pm 6$~meV to $64 \pm 7$~meV.
This can be attributed to temperature-dependent electron--phonon ($\gamma_\mathrm{e-ph}$) or electron--electron ($\gamma_\mathrm{e-e}$) scattering rates that both contribute~\cite{Link1999, Liu2009} to the total relaxation rate $\gamma = \gamma_\mathrm{e-ph} + \gamma_\mathrm{e-e}$.
While the electron--phonon scattering rate depends on the lattice temperature that is unlikely to vary significantly during the pulse duration of $< 100$~fs, the electron--electron scattering rate follows the electron temperature as $\gamma_\mathrm{e-e} \propto \left(k_B T_e\right)^2 + \left(\hbar\omega\right)^2$.
Considering the approximately linear power dependence of $T_e$, the corresponding power dependence of the relaxation rate can be described by a quadratic function $\gamma (P) = \gamma_0 + \alpha P^2$ (red curves, error-weighted fit), with the extracted temperature-independent contribution of $\gamma_0 = 41 \pm 6$~meV.

The described in this work enhancement of the optical nonlinearity in nanotips and nanorods through a strong dipole-forbidden third-order contribution represents a general mechanism for nano-structured media with free carriers where high field gradients can be achieved.
In contrast to conventional approaches for increased nonlinearity in metallic nano-particles relying on extrinsic local field enhancement, this gradient-field effect modifies the microscopic nonlinear susceptibility, and can therefore contribute even in the absence of intrinsic optical nonlinearities of the medium.
Further, it provides enhancement in materials that already possess intrinsic nonlinearities, e.g., in graphene, where the strong third-order susceptibility has received much attention recently~\cite{Glazov2014}, yet with monolayer volume offering small signal levels that can now be increased further through nano-structuring and engineering large field gradients.
With $1/R^{n-1}$ size scaling (Eq.~\ref{eq:Lorentzian}) of the gradient-field term in the $n$-order susceptibility $\chi^{(n)}$, its contribution increases favorably with decreasing sample volume and increasing order of the nonlinearity.
The results of this work not only offer insight into the microscopic mechanisms of the nonlinear optical response on the nanoscale through the hydrodynamic plasma model, but also demonstrate a qualitatively new approach to nonlinear nano-optics opening new avenues from on-chip nonlinear all-optical information processing to coherent multidimensional nano-spectroscopy and -imaging.

%, which makes it useful for applications requiring strong nonlinear optical signals on the nanoscale, such as 2D nano-spectroscopy and all-optical nano-devices.

%\section{Conclusions}
%AB 
%In summary, we demonstrated strong electric dipole-forbidden third-order optical nonlinearity of Au nanotips and nanorods enabled by large plasmonic field gradients. 
%This mechanism leads to the enhancement of the third-order nonlinear susceptibility $> 5$ times that of bulk Au  in the transition from non-degenerate to degenerate FWM for the excitation photon energies around $1.56$~eV.
%The enhancement of the FWM signal is sensitive to the tip apex geometry, and can reach up to 10 for some tips used in our experiment.
%AB
%Our experimental data is described well by a simple model based on the hydrodynamic approach for a free-carrier plasma, which predicts a Lorentzian shape of the FWM efficiency vs. detuning $\delta\omega = \omega_1 - \omega_2$.
%The results presented here provide a new insight into the microscopic mechanism of the nonlinear optical response in metal nanostructures, and suggest applications for on-chip all-optical information processing and coherent multidimensional spectroscopy on the nanoscale.

\section{Acknowledgements}
We acknowledge funding from the National Science Foundation (NSF Grant CHE 1709822).
A.B. and M.R. also acknowledge support from the Air Force Office for Scientific Research through grant FA9550-14-1-0376.
R.U. acknowledges support by a Rubicon Grant of the Netherlands Organization for Scientific Research (NWO).

%\bibliographystyle{naturemag}
%\bibliography{gradientfwm}

\end{document}